\documentclass[aps,prll,twocolumn,superscriptaddress,amsmath,amssymb]{revtex4}

\usepackage{natbib}
\usepackage{hyperref}
\usepackage{graphicx}
\usepackage{dcolumn}
\usepackage{bm}
\usepackage{hyperref}
\usepackage[mathlines]{lineno}
\usepackage[dvipsnames]{xcolor}

\begin{document}
	
\title{CeAu$_{2}$Bi: a new nonsymmorphic antiferromagnetic compound}

\author{M. M. Piva}
\email{mpiva@ifi.unicamp.br}
\affiliation{Instituto de F\'{\i}sica ``Gleb Wataghin'', UNICAMP, 13083-859, Campinas, SP, Brazil}
\affiliation{Los Alamos National Laboratory, Los Alamos, New Mexico 87545, USA}

\author{W. Zhu}
\affiliation{Los Alamos National Laboratory, Los Alamos, New Mexico 87545, USA}
\affiliation{Institute of Natural Sciences, Westlake Institute of Advanced Study and School of Science, Westlake University, Hangzhou, 310024, China}

\author{F. Ronning}
\affiliation{Los Alamos National Laboratory, Los Alamos, New Mexico 87545, USA}

\author{J. D. Thompson}
\affiliation{Los Alamos National Laboratory, Los Alamos, New Mexico 87545, USA}

\author{P. G. Pagliuso}
\affiliation{Instituto de F\'{\i}sica ``Gleb Wataghin'', UNICAMP, 13083-859, Campinas, SP, Brazil}

\author{P. F. S. Rosa}
\affiliation{Los Alamos National Laboratory, Los Alamos, New Mexico 87545, USA}

\date{\today}

\begin{abstract}
	
	Here we report the structural and electronic properties of CeAu$_{2}$Bi, a new heavy-fermion compound crystallizing in a nonsymmorphic hexagonal structure ($P63/mmc$). The Ce$^{3+}$ ions form a triangular lattice which orders antiferromagnetically below $T_{N} = 3.1$~K with a magnetic hard axis along the c-axis. Under applied pressure, $T_{N}$ increases linearly at a rate of $0.07$~K/kbar, indicating that the Ce $f$-electrons are fairly localized. In fact, heat capacity measurements provide an estimate of 150(10) mJ/mol.K$^{2}$ for the Sommerfeld coefficient. The crystal-field scheme obtained from our thermodynamic data points to a ground state with dominantly $|j_{z}=\pm1/2\rangle$ character, which commonly occurs in systems with a hard c-axis. Finally, electronic band structure calculations and symmetry analysis in $k$-space reveal that CeAu$_{2}$Bi hosts symmetry-protected crossings at $k_{z} = \pi$ in the paramagnetic state.
	
\end{abstract}

\maketitle

\section{INTRODUCTION}

Materials discovery is an important part of condensed matter research and is driven by the possibility of finding materials that may host new physical phenomena and enable future technologies. Recently, compounds with topologically protected surface states are being extensively pursued due to their potential applications. Though topology concepts have been extended to condensed matter physics, they remain rather unexplored in 4$f$ Cerium-based materials \cite{CeSbTe, RAlGe, CeSb, KondoIns}. Many of these materials crystallize in nonsymmorphic structures - which are predicted to generate protected band crossings as well as surface states \cite{Parameswaran, Shiozaki, Yang, Zhang}. In addition, Ce-based materials often display emergent properties such as complex magnetism and unconventional superconductivity, and the interplay between topology and emergence could lead to new quantum states of matter. Nonsymmorphic topological insulators were predicted to host surface fermions, which exhibit an \textquotedblleft hourglass"-shaped dispersion \cite{Wang, Alex, Shiozaki, Chang, Wieder}. Recent experiments in KHgSb provide experimental evidence of such fermions \cite{Ma}. Further, nonsymmorphic symmetries may produce band touching, hindering the formation of a gapped insulator. As a result, a nodal semimetal ground state is realized in which band degeneracies are required by symmetry. Hexagonal space groups are notable for their sixfold screw rotation symmetry which leads to multiple band crossings with an \textquotedblleft accordion-like" dispersion \cite{Zhang}. UPt$_{3}$, for instance, crystallizes in the same nonsymmorphic hexagonal space group as KHgSb, $P63/mmc$ (SG 194), and may be a topological superconductor \cite{Mobius}. In fact, hexagonal materials in space groups containing both inversion and glide mirror symmetries are predicted to host symmetry protected band crossings, pinned to high-symmetry plane $k_{z} = \pi$ \cite{Zhang}. Therefore, materials that crystallize in the $P63/mmc$ (SG 194) structure are promising to display non-trivial topology features. 

\begin{figure}[!b]
	\includegraphics[width=0.5\textwidth]{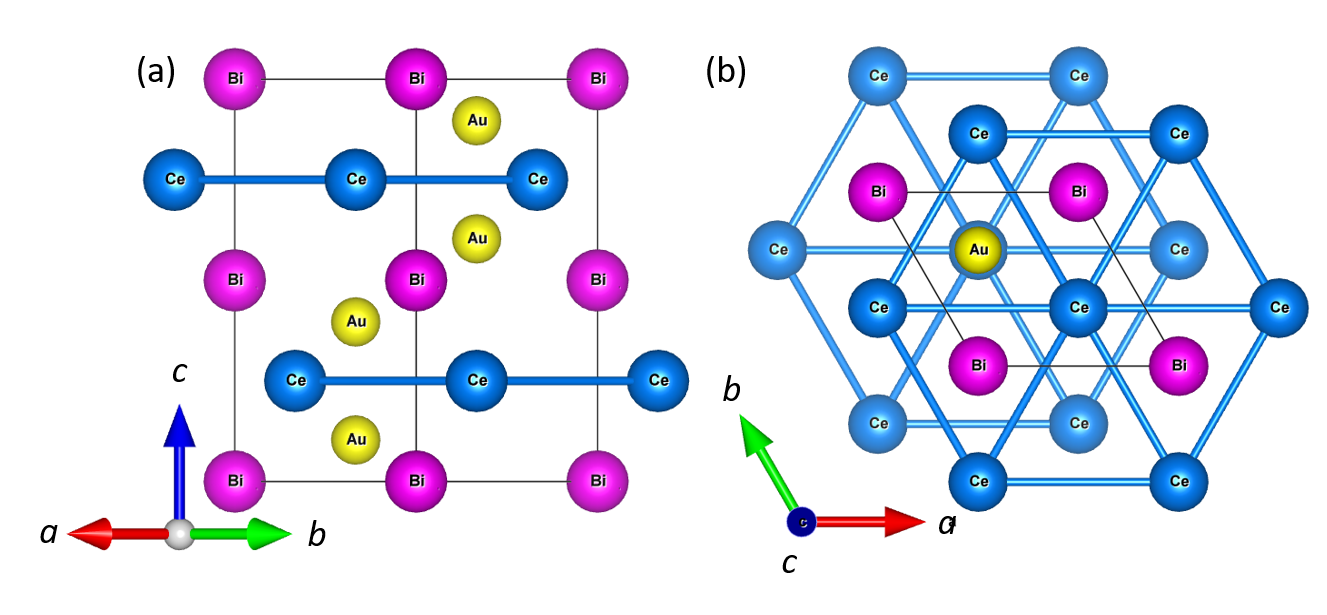}
	\caption{(a) Hexagonal structure of CeAu$_{2}$Bi $a$ = 4.887(1)~\AA \- and $c$ = 9.375(1)~\AA. (b) $c$-axis view of the crystalline structure of CeAu$_{2}$Bi displaying the triangular lattice formed by Ce$^{3+}$ ions.}
	\label{xtal}
\end{figure}   

Here, we report the electronic and structural properties of the new Ce-based compound CeAu$_{2}$Bi that also crystallizes in the $P63/mmc$ hexagonal structure. Single crystals of CeAu$_{2}$Bi and LaAu$_{2}$Bi were grown by the Bi-flux technique as described in the Supplemental Material along with other experimental details \cite{SM}. As we will show, CeAu$_{2}$Bi is a moderately heavy-fermion compound that orders antiferromagnetically below 3.1~K out of a $|j_{z}=\pm1/2\rangle$ crystal-field doublet. Though applied pressure increases the ordering temperature, a field of 4.5~T applied perpendicular to the c-axis is sufficient to suppress long range order, similar to CePd$_{2}$In \cite{Structure, SMock, ABianchi, Muon, Chiao, Weller}. Finally, motivated by predictions of band crossings at high symmetry planes \cite{Zhang}, we also present the calculated band structure of CeAu$_{2}$Bi, which reveals a symmetry protected band crossing at $k_z=\pi$ as well as other trivial bands at $E_{F}$. These electronic structure calculations along with experiments support the possibility that CeAu$_{2}$Bi is a moderately correlated antiferromagnet that may host a non-trivial electronic topology. Our experimental data, however, do not provide evidence for the predicted topological states as trivial bands overwhelm the electronic response at $E_{F}$.

\section{RESULTS}

\begin{figure}[!t]
	\includegraphics[width=0.5\textwidth]{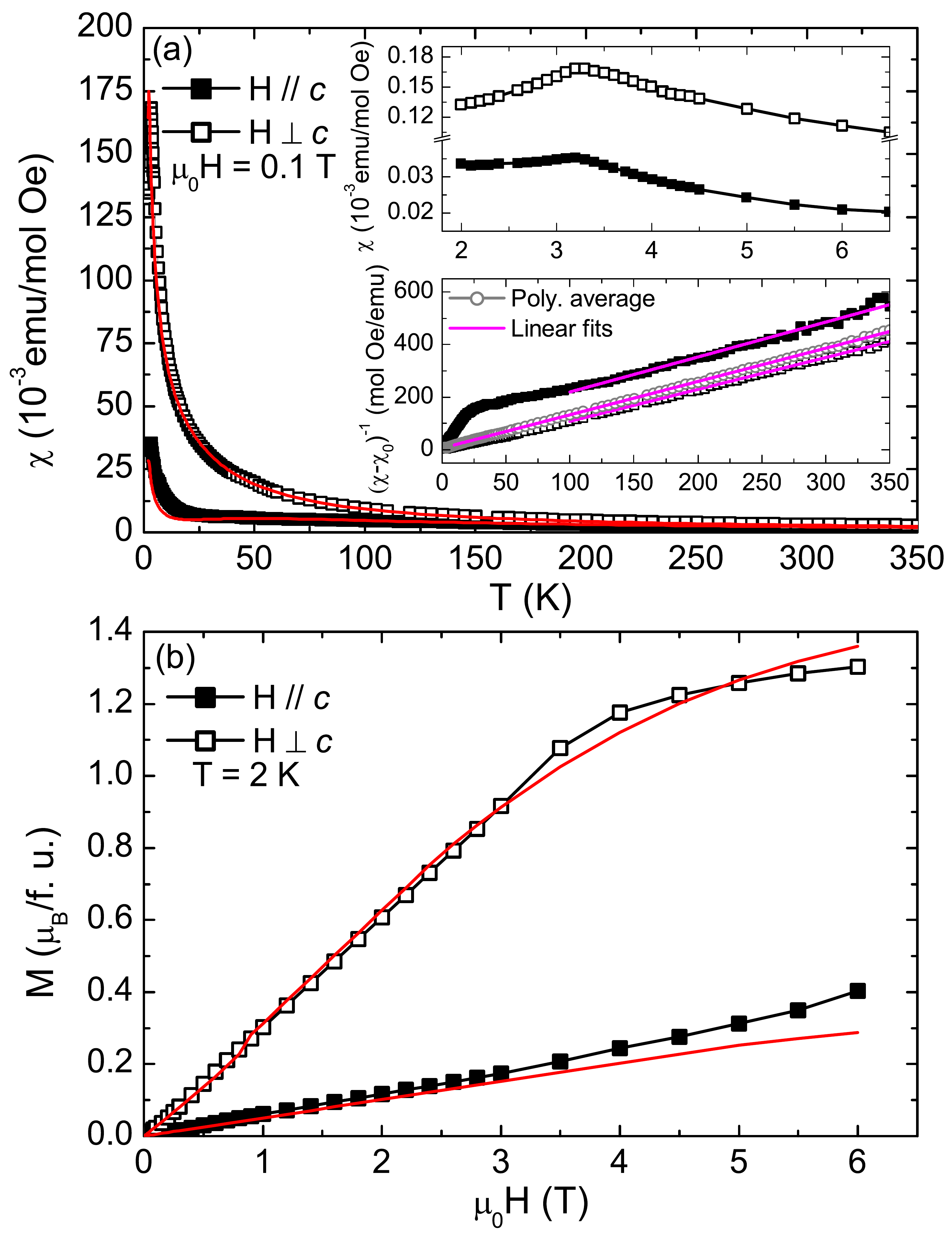}
	\caption{(a) $\chi(T)$ at $\mu_{0}H = 0.1$~T. The top inset shows a zoomed-in view of $\chi(T)$ in the low-temperature range. The bottom insets shows $(\chi -\chi_{0})^{-1}$ as a function of temperature. (b) $M$ as a function of applied magnetic field. The solid red lines are fits using a CEF mean field model.}
	\label{chi}
\end{figure}

Figure~\ref{xtal}(a) shows the hexagonal crystal structure shared by CeAu$_{2}$Bi and LaAu$_{2}$Bi. In this structure, Ce/La planes are separated by AuBi layers. Figure~\ref{xtal}(b) presents the triangular lattice formed by Ce$^{3+}$ ions. The crystallographic data of both compounds are summarized in Table I in the Supplemental Material \cite{SM}. Both compounds crystallize in the hexagonal $P63/mmc$ nonsymmorphic structure, although CeAu$_{2}$Bi has a slightly smaller unit cell than LaAu$_{2}$Bi, as expected from the lanthanide series contraction. There is no detectable second phase in single crystal diffraction, though a small ($< 5$~\%) amount of impurity phases cannot be ruled out, and indeed there is evidence for a Au$_{2}$Bi impurity phase in both CeAu$_{2}$Bi and LaAu$_{2}$Bi in resistivity measurements, see Supplemental Material for more details \cite{SM}. 

\begin{figure}[!t]
	\includegraphics[width=0.5\textwidth]{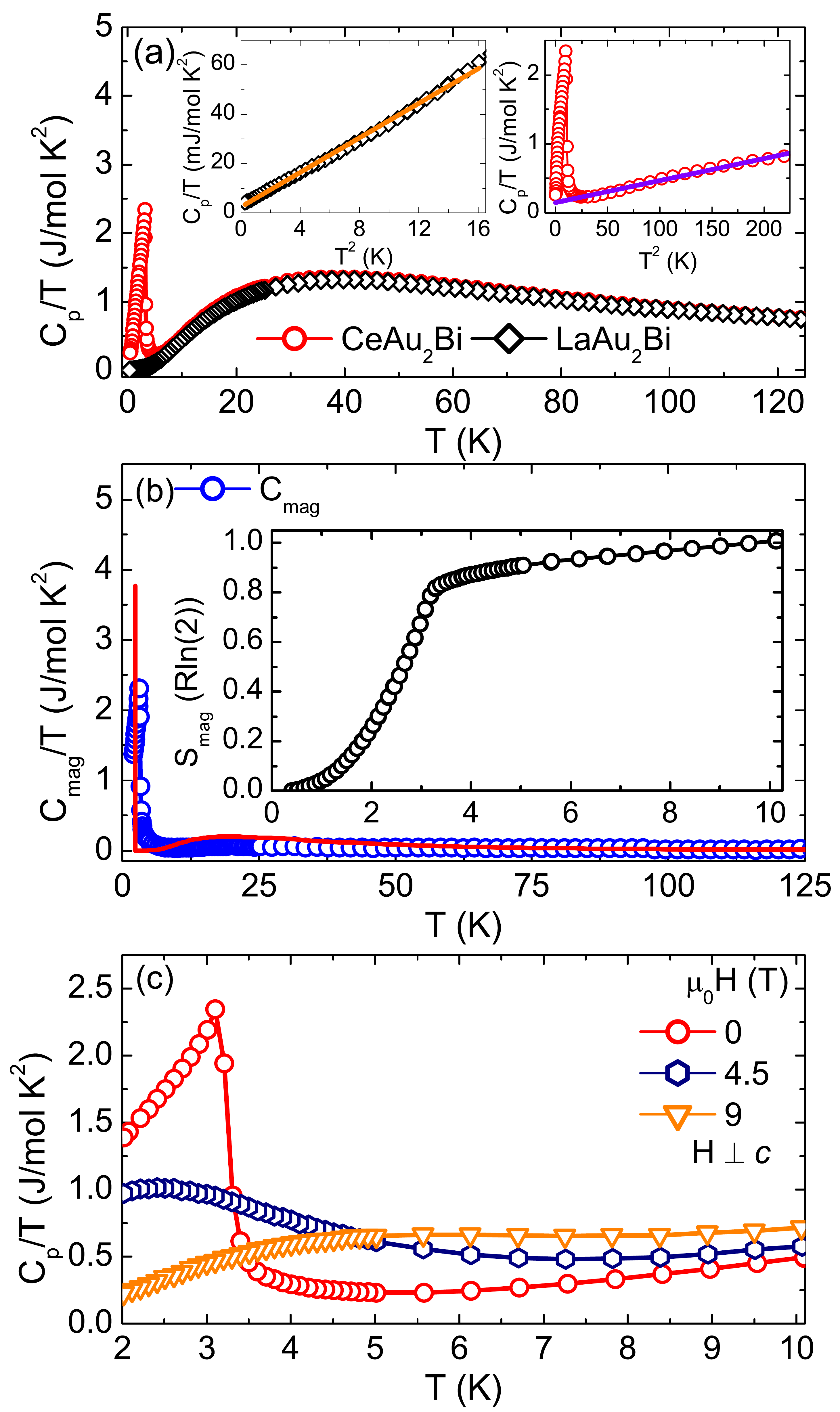}
	\caption{(a) Specific heat measurements for CeAu$_{2}$Bi and LaAu$_{2}$Bi. The insets show $C_{p}/T$ as a function of $T^{2}$. The solid line is an extrapolation of the fit used to extract the Sommerfeld coefficient. (b) Magnetic specific heat and entropy of CeAu$_{2}$Bi. The solid red line is a fit using a CEF mean field model. (c) Specific heat behavior as a function of temperature of CeAu$_{2}$Bi for several applied magnetic fields.}
	\label{cp}
\end{figure}

Figure~\ref{chi}(a) shows the magnetic susceptibility ($\chi$) as a function of temperature at $\mu_{0}H = 0.1$~T applied parallel and perpendicular to the $c$-axis of CeAu$_{2}$Bi. These measurements find that the $c$-axis is the hard axis and that Ce$^{3+}$ moments order antiferromagnetically at $T_{N} = 3.1$~K, as we can clearly see in the top inset of Fig.~\ref{chi}(a). The bottom inset shows the inverse of $\chi$ after subtraction of a Pauli contribution ($\chi_{0} \approx 0.4$~10$^{-3}$emu/(mol.Oe)) for $H \ || \ c$, $H \perp c$ and for the polycrystalline average ($\frac{1}{3}\chi_{||c} + \frac{2}{3}\chi_{\perp c}$). By performing linear fits in the high-temperature range, we extract the following Ce$^{3+}$ effective moments: $\mu_{eff} = 2.57(1)$~$\mu_{B}$, when $H \perp c$; $\mu_{eff} = 2.46(2)$~$\mu_{B}$ for $H \ || \ c$ and $\mu_{eff} = 2.52(1)$~$\mu_{B}$ for polycrystalline average. These values are close to the calculated value of 2.54~$\mu_{B}$ for a free Ce$^{3+}$ ion. From these fits, we find Curie-Weiss temperatures of $\theta_{CW} = 10(1)$~K with $H \perp c$, $\theta_{CW} = -67(4)$~K with $H \ || \ c$, and $\theta_{CW} = -5.5(2)$~K from the polycrystalline average. Assuming that CEF (crystalline electrical field) effects dominate $\theta_{CW}$, we consider the polycrystalline average to calculate the frustration parameter ($f = |\theta_{CW}|/T_{N}$) of around 1.8, which indicates that only weak geometrical frustration is present in CeAu$_{2}$Bi. At low temperatures, one can see a deviation from the Curie-Weiss behavior when $H \ || \ c$, which is likely associated with crystal field effects, as we will discuss below.

Figure~\ref{chi}(b) displays the magnetization ($M$) behavior as a function of applied magnetic field at 2~K. For $H \perp c$, $M$ starts to saturate at 3.5~T, pointing to a fully polarized spin state at higher fields. In fact, the saturated moment of $\approx$~1.3~$\mu_{B}$ is expected for a $|j_z = \pm 1/2>$ doublet ground state in hexagonal symmetry, in agreement with Ref.~\cite{REHex} and similar to CeAgSb$_{2}$ \cite{CeAgSb2}. For $H \ || \ c$ the magnetization increases monotonically with field. 

Figure~\ref{cp}(a) displays the specific heat divided by the temperature ($C_{p}/T$) for both CeAu$_{2}$Bi and LaAu$_{2}$Bi as a function of temperature. CeAu$_{2}$Bi displays a lambda-type peak at 3.1~K which characterizes $T_{N}$ and is in agreement with $\chi(T)$ results. Moreover, the insets of Figure~\ref{cp}(a) plot $C_{p}/T$ as a function of $T^{2}$ for both compounds. 
The solid line in the left inset is a linear fit that gives the Sommerfeld coefficient ($\gamma$) of 2.4(2)~mJ/(mol.K$^{2}$) for LaAu$_{2}$Bi. An extrapolation to $T = 0$~K of a linear fit between 5~K and 15~K of $C_{p}/T$ vs $T^{2}$ above $T_{N}$  (right inset) gives an estimated $\gamma = 150(10)$~mJ/(mol.K$^{2}$) for CeAu$_{2}$Bi, that characterizes it as a moderately heavy-fermion compound. By subtracting the phonon contribution for the specific heat of CeAu$_{2}$Bi assumed to be given by the specific heat of its non-magnetic analog LaAu$_{2}$Bi, we obtain the magnetic specific heat, that is displayed in Fig.~\ref{cp}(b). The integral of $C_{mag}/T$ vs $T$ gives the magnetic entropy of CeAu$_{2}$Bi, which is plotted in the inset of Fig.~\ref{cp}(b) and reaches 80~\% of the doublet entropy $Rln(2)$ at $T_{N}$. The presence of magnetic entropy above $T_{N}$ could arise from a partial Kondo effect but also from short range magnetic correlations. $C_{p}/T$ as a function of temperature for CeAu$_{2}$Bi at different applied magnetic fields is displayed in Fig.~\ref{cp}(c). Similar to CePd$_{2}$In \cite{Chiao}, $T_{N}$ is suppressed as a function of increasing field, being fully suppressed at 4.5~T and in agreement with our magnetization data.

\begin{figure}[!t]
	\includegraphics[width=0.5\textwidth]{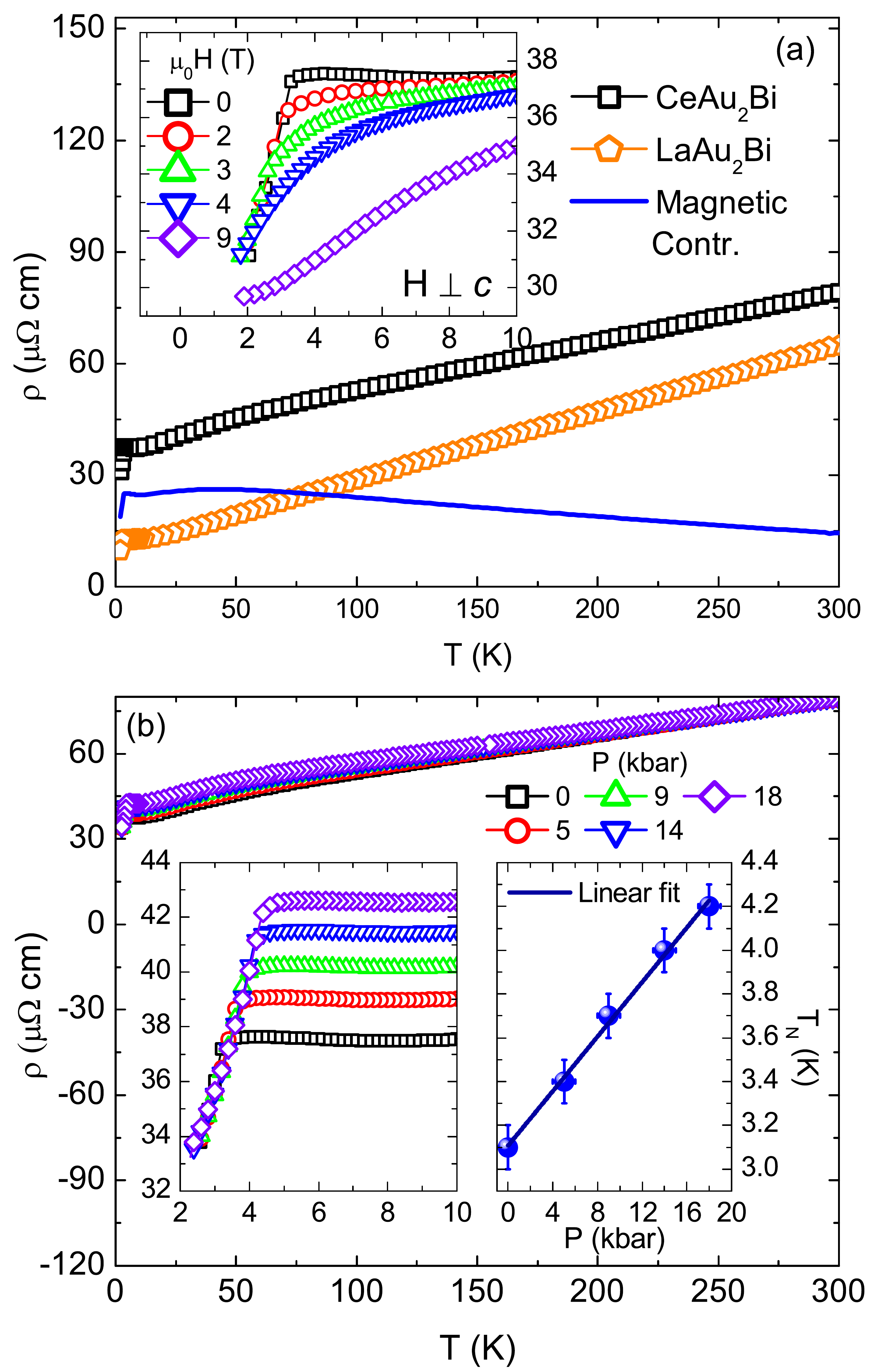}
	\caption{(a) $\rho(T)$ for CeAu$_{2}$Bi and LaAu$_{2}$Bi. The inset shows the low-temperature behavior of the CeAu$_{2}$Bi resistivity for several applied magnetic fields. (b) $\rho(T)$ of CeAu$_{2}$Bi for different applied pressures. The insets show the low-temperature range of the resistivity for several applied pressures and the temperature-pressure phase diagram, respectively.}
	\label{rho}
\end{figure}  

\begin{figure*}[!t]
	\includegraphics[width=0.8\textwidth]{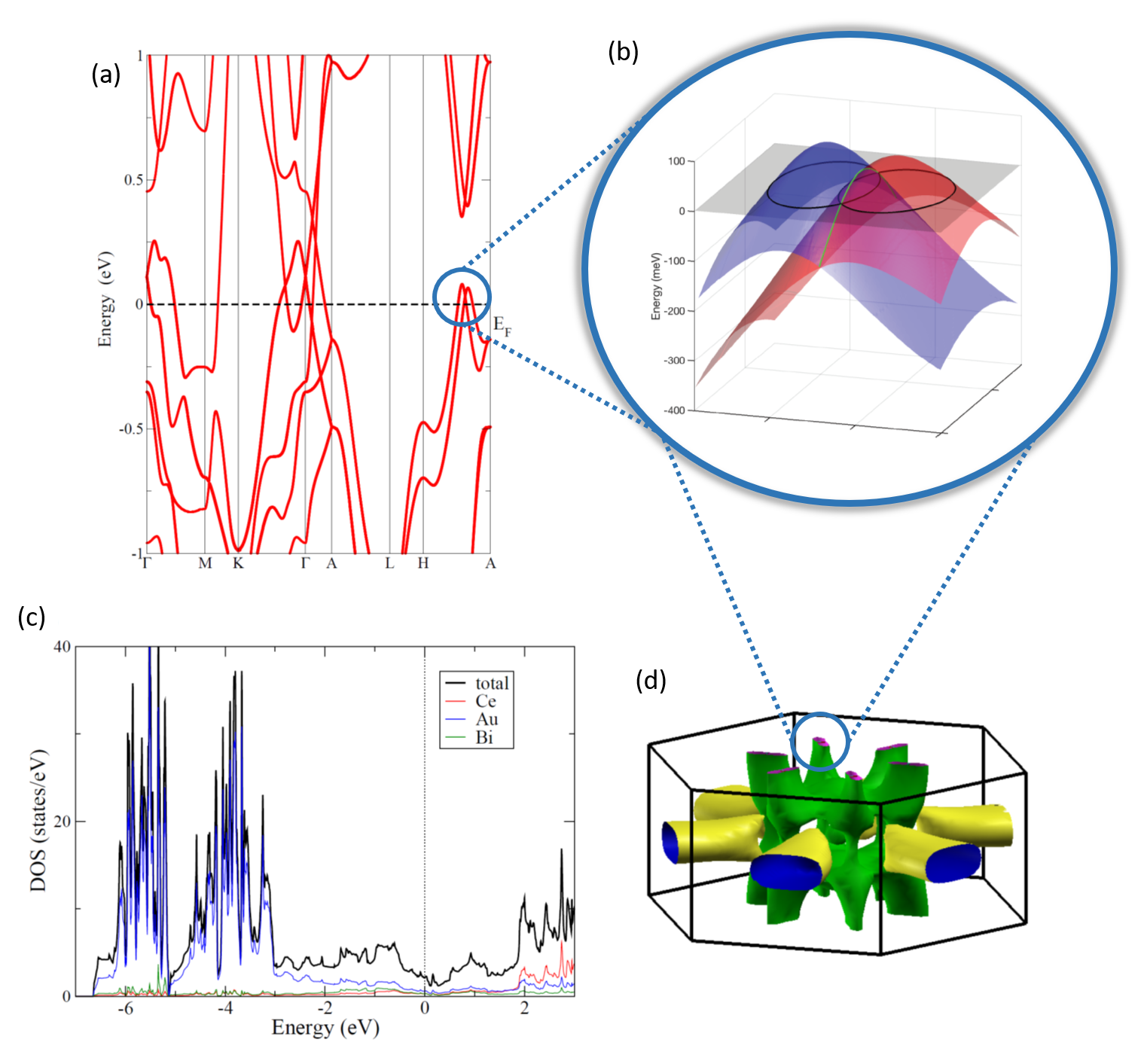}
	\caption{(a) Electronic band structure of CeAu$_{2}$Bi. (b) Zoom in view of the electronic structure for $k_{z} = \pi$. Two bands with opposite $R-{z}$ eigenvalues (red and blue) intersect forming a line (green), which intersects the chemical potential (gray plane) at the two points where the Fermi surfaces of the individual bands (black circles) intersect (see text for details). (c) Density of states and (d) Fermi surface of CeAu$_{2}$Bi.}
	\label{calc}
\end{figure*}

The solid red lines in the main panels of Figs.~\ref{chi}(a) and (b) and in Fig.~\ref{cp}(b) are fits to data using a CEF mean field model considering anisotropic nearest-neighbor interactions and the hexagonal CEF Hamiltonian: $\mathcal{H} = z_{i}J_{i} J \cdot  \langle J\rangle + B^{0}_{2}O^{0}_{2} + B^{0}_{4}O^{0}_{4} + B^{0}_{6}O^{0}_{6} + B^{6}_{6}O^{6}_{6}$. $z_{i}J_{i}$ represents interactions ($i = $AFM, FM) between nearest neighbors that mimic the RKKY interaction, $B^{m}_{n}$ are the CEF parameters and the $O^{m}_{n}$ are the Steven's operators, similar to the approach used in Refs.~\cite{CEFfits, Ali}. By simultaneously performing fits to $\chi(T)$, $M(H)$ and $C_{p}(T)$ data, we extract the CEF scheme and two RKKY parameters for this compound. For the RKKY parameters we obtain $z_{AFM}J_{AFM} = 0.78$~K and $z_{FM}J_{FM} = -0.20$~K. We note that the extracted $T_{N} \approx 2.3$~K is slightly smaller than the experimental value, which could be due to additional exchange interactions not captured by our simple mean field model. For the CEF parameters we obtained the following values: $B^{0}_{2} \approx 7.28$~K, $B^{0}_{4} \approx -0.04$~K, $B^{0}_{6} \approx 0.03$~K and $B^{6}_{6} \approx -0.01$~K. These parameters imply a ground state composed of a $\Gamma_{6}$ doublet ($|\pm1/2\rangle$), a first excited state $|\pm3/2\rangle$ doublet at 60~K and a second excited state $|\pm5/2\rangle$ doublet at 130~K. We note that the CEF parameters acquired from fits of macroscopic data may not be fully accurate or unique. Therefore, to unambiguously determine the CEF scheme, inelastic neutron scattering and X-ray absorption experiments should be performed. 

Figure~\ref{rho}(a) displays the electrical resistivity ($\rho$) of CeAu$_{2}$Bi, LaAu$_{2}$Bi and the magnetic resistivity ($\rho$(CeAu$_{2}$Bi) - $\rho$(LaAu$_{2}$Bi)). CeAu$_{2}$Bi displays metallic behavior but with a rather poor relative resistance ratio $R(300K)/R(2K) \approx 2.5$. Presently, we do not know if this ratio is intrinsic or possibly due to weak site disorder. Nevertheless, there is a well-defined kink around 3.1~K that reflects the loss of spin-disorder scattering below $T_{N}$. Near 50~K there is a broad hump in magnetic resistivity that is most likely associated with the depopulation of the first excited crystal field state. In the inset of Fig.~\ref{rho}(a) we plot the field dependence of $\rho(T)$ with $H \perp c$ near $T_{N}$. For fields higher than 3~T $T_{N}$ cannot be identified anymore above 1.8~K, in agreement with the specific heat measurements and similar to CePd$_{2}$In \cite{Chiao}. Moreover, in the Fermi-liquid state at 9~T  where $\rho$ increases as $T^{2}$, the coefficient of the $T^{2}$ term is $A = 0.10(1)$~$\mu \Omega$.cm/K$^{2}$. This value of A implies a $\gamma$ of $100(20)$~mJ/(mol.K$^{2}$) considering the Kadowaki-Woods ratio \cite{KW} and is in good agreement with the coefficient extracted from specific heat measurements. LaAu$_{2}$Bi also displays metallic behavior at high temperatures, whereas at low temperatures a downturn is observed near 2~K, likely due to the superconducting transition of the extrinsic binary Au$_{2}$Bi at 1.8~K \cite{SM, Au2Bi}, as no evidence of superconductivity was found in specific heat measurements. Figure~\ref{rho}(b) presents $\rho(T)$ of CeAu$_{2}$Bi at several applied pressures. At high temperatures, CeAu$_{2}$Bi displays metallic behavior for all pressures. The left inset in Fig.~~\ref{rho}(b) displays the pressure dependence of the $\rho(T)$, and the right inset shows the linear increase of $T_{N}$ with pressure at a rate of 0.069(1)~K/kbar. The increase of $T_{N}$ with pressure within the Doniach diagram further supports the localized nature of the $f$-electrons in CeAu$_{2}$Bi.

To explore the possibility that CeAu$_{2}$Bi might present a non-trivial electronic structure and to place experimental results in the broader context of its global electronic structure, we perform \textit{ab initio} calculations based on density functional theory \cite{Hohenberg, Blaha, Perdew}. We compute the band structure of CeAu$_2$Bi with spin-orbit
coupling and considering the Ce 4$f$ electron localized in the core, as illustrated in Fig.~\ref{calc}(a). Because the magnetic structure is unknown we also consider the non-magnetic solution relevant for the paramagnetic state above $T_{N}$, as well as the non-magnetic La analog. Let us consider each of the high-symmetry lines.
All bands in Fig.~\ref{calc}(a) are at least doubly degenerate because the crystal structure has inversion symmetry $\hat P$ and (neglecting magnetic order) time-reversal symmetry $\hat T$, which ensures that the bands are Kramers degenerate for all $\mathbf k$ \cite{Young2015}. At any wavevector along the $k_z=0$ plane, bands avoid crossing at all momenta, whereas in the $k_z=\pi$ plane, the band structure shows crossings along the $H-A$ line. 
Symmetry analysis indicates that this crossing is attributed to the nonsymmorphic symmetry $R_{z}$:$\,(x,y,z)\rightarrow(x,y,-z+c/2),(s_x,s_y,s_z)\rightarrow(-s_x,-s_y,s_z)$, 
where $R_z$ is a symmetry combining a twofold screw axis along $z-$direction and inversion symmetry. Importantly, we find that $ R_z*(P*T)=e^{-ik_z}(P*T)*R_z $. On the mirror invariant planes $k_z = 0$ and $\pm\pi$, the bands can be labeled by their $R_z$ eigenvalues $\pm i$ because $R_z^2=-1$. 
This means that the degenerate bands on the $k_z=0$ plane have opposite $R_z$ eigenvalues $(i,-i)$, and two sets of such doublet bands generally anticross, i.e., bands with the same $R_z$ eigenvalues avoid crossing.
At $k_{z} = \pi$, however, the degenerate bands have the same $R_{z}$ eigenvalue $\pm(i,i)$. In this case, two doublet bands with opposite $R_{z}$ eigenvalues may cross, creating a symmetry protected four-band crossing line \cite{Fang2015,Fang2016}. In Fig.~\ref{calc}(b), we show a zoom-in figure revealing the crossing bands whose intersection passes through the chemical potential. One can see that the protected band crossings create a line (green solid line) that extends in a direction perpendicular to H-A. This line does form a ring about the A point but is in fact intersected by the fourfold band degeneracy along A-L and symmetry equivalent directions. Away from the $k_{z} = π$ plane, $R_{z}$ is no longer a good symmetry and the bands hybridize, opening a gap. Although it is exciting to find a symmetry protected crossing intersecting the Fermi energy in an $f$-electron system, we note that there exist multiple trivial Fermi surface sheets (Fig.~\ref{calc}(d)), which likely dominate the transport and thermodynamic properties of this compound. Furthermore, our DFT calculations do not consider the presence of heavy electrons and antiferromagnetic order. The former would require DFT+DMFT calculations to take correlations into account more precisely. The latter would require the determination of the magnetic structure, which is beyond the scope of the present work.

At the $k_z=\pi$ plane, the band structure shows four-fold degeneracy along the $A-L$ line. 
This higher level degeneracy results from the mirror symmetry $M_y:(x,y,z)\mapsto(x,-y,z)$. The implication is that the high symmetry path $k_y=0$ at $k_z=\pi$ is just the intersecting line of the invariant plane $k_y=0$ of $M_y$ and the invariant plane $k_z=\pi$ of $R_z$. On this invariant line, action of $\mathbf{M}_y$ will transform the eigenstate to a partner of an opposite $R_z$-label, which then relates the two Krammer pairs and ensures a four-fold degeneracy. In Fig.~\ref{calc}(c) we show the calculated density of states for CeAu$_{2}$Bi. Extracting the Sommerfeld coefficient from this density of states yields a $\gamma$ of 2.6(3)~mJ/(mol.K$^{2}$). This value is in good agreement with the $\gamma$ of 2.4(2)~mJ/(mol.K$^{2}$) experimentally found for LaAu$_{2}$Bi, i.e. the 4$f$ localized limit of CeAu$_{2}$Bi. Finally, Fig.~\ref{calc}(d) displays the Fermi surface of CeAu$_{2}$Bi, which is six fold symmetric and presents 2D-like cylindrical pockets, along with 3D-like pockets.  

\section{CONCLUSIONS}

In summary,  we report the electronic and structural properties of the new heavy-fermion compound CeAu$_{2}$Bi and of its non-magnetic analog LaAu$_{2}$Bi, which crystallize in the nonsymmorphic $P63/mmc$ hexagonal structure. At high temperatures, both samples show metallic behavior. At low temperatures, CeAu$_{2}$Bi orders antiferromagnetically below 3.1~K with a magnetic hard axis along the $c$-axis. The antiferromagnetic transition recovers 80~\% of $Rln(2)$  indicating that the Ce$^{3+}$ local moments are fairly localized in this compound. Specific heat measurements and the Kadawoki-Woods ratio estimate a Sommerfeld coefficient of 120(40)~mJ/(mol.K$^{2}$) in the paramagnetic state, placing CeAu$_{2}$Bi as a moderately heavy compound. Furthermore, by performing fits of $\chi(T)$, $M(H)$ and $C_{p}(T)$ using a CEF mean field model, we could extract two competing exchange interactions, $z_{AFM}J_{AFM} = 0.8$~K and $z_{FM}J_{FM} = -0.2$~K and a $\Gamma_{6}$ ($j_{z}=|\pm1/2\rangle$) ground state. Moreover, electrical resistivity experiments under pressure revealed an increase of $T_{N}$ as a function of pressure reaching a maximum of 4.2~K at 18~kbar. Neutron diffraction experiments would be useful to solve the magnetic structure and to probe the CEF scheme of this new compound. Also, experiments at higher pressures are needed to investigate whether $T_{N}$ can be suppressed and a quantum critical point can be achieved in CeAu$_{2}$Bi. Finally, to shed new light on the electronic energy levels of CeAu$_{2}$Bi, we performed \textit{ab initio} calculations based on the density functional theory. These calculations show that at the $k_{z}=0$ plane, the band structure has twofold degeneracy and bands avoid crossing at all momenta. At $k_z=\pi$, the band structure is four-fold degenerate along A-L line and possesses multiple band crossings elsewhere in the $k_{z} = \pi$ plane, one of which intersects the Fermi energy. However, no experimental evidence of band crossings were observed in our measurements. In this regard, ARPES experiments would be useful to directly probe the electronic bands, search for band crossings and to determine the influence of magnetism on such topological band crossings.       

\begin{acknowledgments}
	
	We would like to acknowledge N. Harrison and M. Rahn for useful discussions. Sample synthesis and crystal structure determination were supported by the DOE BES ``Science of 100 Tesla" project. Physical property measurements and band structure calculations were supported by the DOE BES ``Quantum Fluctuations in Narrow-Band Systems" project. M.M.P. acknowledges support from the Sao Paulo Research Foundation (FAPESP) grants 2015/15665-3, 2017/25269-3, 2017/10581-1, CAPES and CNPq, Brazil. W.Z. thanks the startup funding from Westlake University. Scanning electron microscope and energy dispersive X-ray measurements were performed at the Center for Integrated Nanotechnologies, an Office of Science User Facility operated for the U.S. Department of Energy (DOE) Office of Science. 
	
\end{acknowledgments}

\bibliography{basename of .bib file}

\end{document}


\title{Supporting information for: CeAu$_{2}$Bi: a new nonsymmorphic antiferromagnetic compound}

\author{M. M. Piva}
\affiliation{Instituto de F\'{\i}sica ``Gleb Wataghin'', UNICAMP, 13083-859, Campinas, SP, Brazil}
\affiliation{Los Alamos National Laboratory, Los Alamos, New Mexico 87545, USA}

\author{W. Zhu}
\affiliation{Los Alamos National Laboratory, Los Alamos, New Mexico 87545, USA}
\affiliation{Institute of Natural Sciences, Westlake Institute of Advanced Study and School of Science, Westlake University, Hangzhou, 310024, China}

\author{F. Ronning}
\affiliation{Los Alamos National Laboratory, Los Alamos, New Mexico 87545, USA}

\author{J. D. Thompson}
\affiliation{Los Alamos National Laboratory, Los Alamos, New Mexico 87545, USA}

\author{P. G. Pagliuso}
\affiliation{Instituto de F\'{\i}sica ``Gleb Wataghin'', UNICAMP, 13083-859, Campinas, SP, Brazil}

\author{P. F. S. Rosa}
\affiliation{Los Alamos National Laboratory, Los Alamos, New Mexico 87545, USA}

\date{\today}



\maketitle

\section{EXPERIMENTAL DETAILS}

Single crystals of CeAu$_{2}$Bi and LaAu$_{2}$Bi were grown by the Bi-flux technique with starting composition Ce/La:Au:Bi = 1:4:10. The reagents were put in an alumina crucible placed inside an evacuated quartz tube. For CeAu$_{2}$Bi, the tube was heated to 1050$^{\circ}$C at 100$^{\circ}$C/h and was kept there for 8 hours. The solution was then cooled down to 630$^{\circ}$C at 8$^{\circ}$C/h, and then to 500$^{\circ}$C at 150$^{\circ}$C/h, where it was annealed for a week. The excess of Bi was removed by spinning the tube in a centrifuge. For LaAu$_{2}$Bi, the mixture of elements was heated to 1050$^{\circ}$C at 100$^{\circ}$C/h and was kept there for 8 hours. Then it was cooled to 500$^{\circ}$C at 3$^{\circ}$C/h. The Bi excess was also removed by centrifuging. Single crystals of both batches presented a Au$_{2}$Bi impurity phase. The crystallographic structure was verified by single-crystal diffraction at room temperature using Mo radiation in a commercial diffractometer. In addition, the synthesized phase was checked by energy dispersive X-ray spectroscopy (EDX). Magnetization measurements were performed in a commercial SQUID-based magnetometer. Specific heat measurements were made using the thermal relaxation technique in a commercial measurement system. A four-probe configuration was used in the electrical resistivity experiments performed using a low-frequency AC bridge. Pressures to 18~kbar were achieved using a self-contained double-layer piston-cylinder-type Cu-Be pressure cell with an inner-cylinder of hardened NiCrAl. Daphne oil was used as the pressure transmitting medium and lead as a manometer. Electronic structure calculations within density functional theory \cite{Hohenberg} were performed with the WIEN2k package \cite{Blaha}. The Perdew-Burke-Ernzerhof exchange-correlation potential based on the generalized gradient approximation was used \cite{Perdew}, and spin-orbit interactions were included through a second variational method. On the basis of the heat capacity measurements, one $f$-electron from cerium was localized in the core.
	

\pagebreak

\section{Crystallographic data}

Crystallographic data of CeAu$_{2}$Bi and LaAu$_{2}$Bi compounds are summarized in Table I.

\begin{table}[h]
 \begin{center}
 \caption{Crystallographic data of CeAu$_{2}$Bi and LaAu$_{2}$Bi determined by single crystal X-ray diffraction.}
 \label{xray}
 \begin{tabular}{c c c}
 \hline
   Crystal system            &  $\hspace{2.7cm}$ &            Hexagonal    \\
   Space group               &  $\hspace{2.7cm}$ &       	   $P63/mmc$ (194) \\
   Temperature               &  $\hspace{2.7cm}$ &              296(2) K    \\
   Wavelength                &  $\hspace{2.7cm}$ &             0.71073 \AA  \\
   $\mu$ (Mo K$\alpha$) (cm$^{-1}$) & \- &      16.348         \\																																																	
		\hline
		& CeAu$_{2}$Bi & LaAu$_{2}$Bi \\
		\hline
   2$\theta_{min}$           &                4.35$^{o}$   & 4.33$^{o}$    \\
   2$\theta_{max}$           &                34.86$^{o}$  & 38.59$^{o}$  \\
       
	 Index ranges              &            & \\    
	      h                    &                -7 -- 7 & -7 -- 8      \\
	      k                    &                -7 -- 6   & -8 -- 8       \\
				l										 &								-15 -- 14 &	-16 -- 16			 \\				 
		                        
		Formula Weigth & 743.03 & 741.82 \\
		                        
		Lattice Parameters & & \\
		a(\AA) & 4.887(1) & 4.906(1)\\
		c(\AA) & 9.375(1) & 9.421(2)\\
		V$_{cell}$(\AA$^{3}$) & 193.88(5) & 196.38(9)\\
		                        
		Atomic positions & & \\
		Ce/La & 1/3 2/3 1/4 & 1/3 2/3 1/4 \\
		Au & 1/3 2/3 0.60324(7) & 1/3 2/3 0.60341(11) \\
		Bi & 0 0 0 & 0 0 0 \\
                            
	Refinement Quality & & \\													
	F$^{2}$(\%)            &	6.70 & 8.68 	                \\
  $R(F)$(\%)$^{a}$      &     2.55 & 3.81     \\
  $R_{w}(F_{o}^{2})(\%)^{b}$                &   6.75 & 10.34                 \\	
	\hline
	\end{tabular}
	
	\begin{tabular}{c c}
	$^{a}R(F)$ = $\sum\mid\mid F_{o}\mid -\mid F_{c}\mid\mid/\sum\mid F{_o}\mid$, for $F_{o}^{2}$ $>$ 2$\sigma (F_{o}^{2})$ & \\
   $^{b}R_{w}(F_{o}^{2})$	= $[\sum[w(F_{o}^{2}-F_{c}^{2})^{2}]/\sum wF_{o}^{4}]^{1/2}$ & \\ 
 \end{tabular} 
 \end{center}
 \end{table}


\pagebreak

\section{EDX results}

\begin{figure}[!ht]
\includegraphics[width=0.7\textwidth]{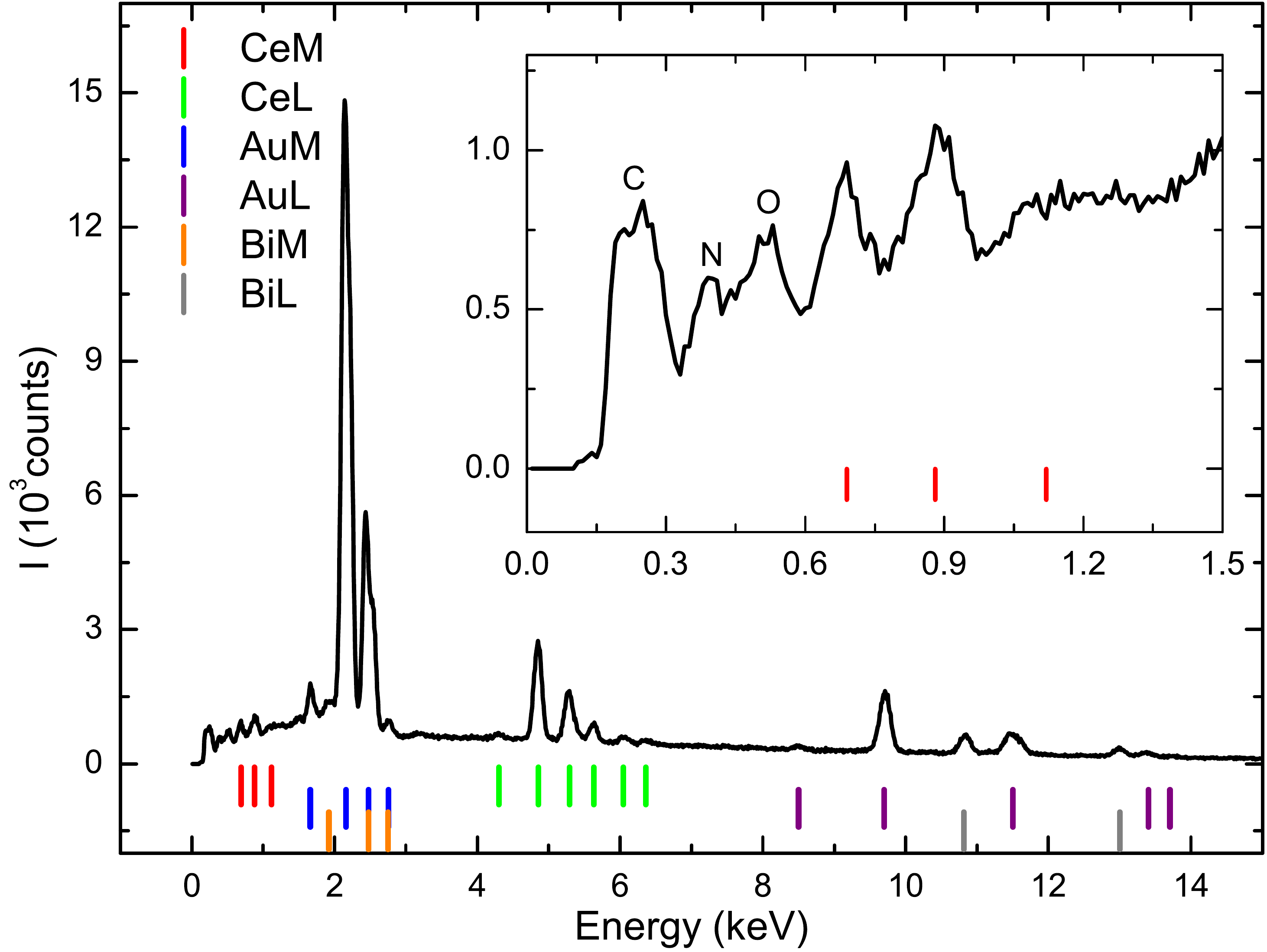}
\caption{EDX spectra for CeAu$_{2}$Bi}
\label{EDX}
\end{figure}

Figure~\ref{EDX} presents the detailed EDX spectra for CeAu$_{2}$Bi.

\pagebreak

\section{Low temperature resistivity}

\begin{figure}[!h]
\includegraphics[width=0.7\textwidth]{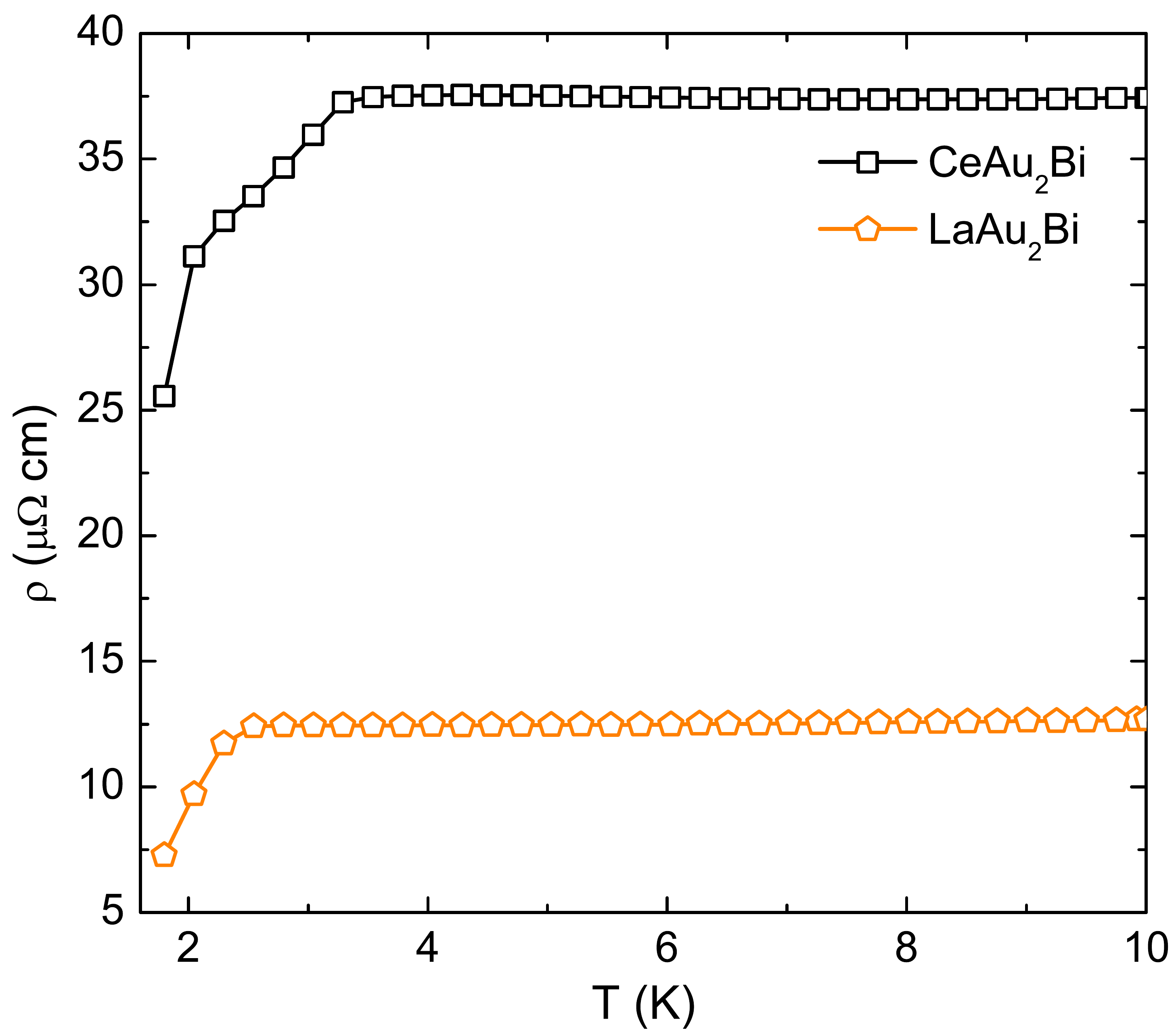}
\caption{Electrical resistivity for CeAu$_{2}$Bi and LaAu$_{2}$Bi at low temperatures ($T < 10$~K).}
\label{lowT}
\end{figure}

Figure~\ref{lowT} presents the electrical resistivity ($\rho$) for CeAu$_{2}$Bi and LaAu$_{2}$Bi at low temperatures ($T < 10$~K). Both samples present a downturn close to 2~K, which may be associated to an Au$_{2}$Bi impurity phase \cite{Au2Bi}. CeAu$_{2}$Bi also displays a kink at $T_{N} = 3.1$~K, due to the loss of spin-scattering.

\bibliography{basename of .bib file}